\documentstyle{espart}

\begin{document}
\begin{frontmatter}
\hfill{IFUP-TH 12/94}

\title{The SU(3) deconfining phase transition with Symanzik action}

\author[IFUP]{\underline{G.~Cella}\thanksref{MURST}},
\author[INFN]{G.~Curci},
\author[INFN]{A.~Vicer\'{e}\thanksref{INFNt}},
\author[IFUP]{B.~Vigna}
\address[IFUP]{Dipartimento di Fisica  dell'Universit\`{a} di Pisa,
Piazza Torricelli 2, I-56126 Pisa, ITALY, e-mail {\tt
cella@sun10.difi.unipi.it}}
\address[INFN]{Istituto Nazionale di Fisica Nucleare, Piazza
Torricelli 2, I-56126 Pisa, ITALY, e-mail {\tt
vicere@sun10.difi.unipi.it}}
\thanks[MURST]{Partially supported by M.U.R.S.T., Italy}
\thanks[INFNt]{Partially supported by I.N.F.N., Italy}

\begin{abstract}
  We report on the determination of the deconfining temperature in
  $SU(3)$ pure gauge theory, using the Symanzik tree level improved
  action, on lattices of size $3\times 12^3,\,4\times 16^3,\,5\times
  20^3,\,6\times 24^3$. We find that the asymptotic scaling violation
  pattern is similar to the one observed using the Wilson action. We
  conclude that the irrelevant operators do not affect, in the range
  of couplings considered, the lattice $\beta$ function. An analysis
  based on an effective coupling formulation shows an apparent
  improvement.
\end{abstract}
\end{frontmatter}

\section{Introduction}

In an earlier work~\cite{cella:curci:tripiccio:vicere} we have reported
on a high statistics investigation of the effect of the Symanzik tree
improvement~\cite{symanzik}, based on the study of the asymptotic
scaling properties of the deconfinement transition in the $SU(2)$ pure
gauge theory.  In this letter we extend the analysis to the
deconfinement transition in $SU(3)$ theory.  Our main purpose is a
better understanding of the effect of the perturbative improvement in
the context of $4$ dimensional gauge theories.  In this respect the
deconfinement temperature $T_c$ is an observable particularly well
suited to give unambiguous answers. As early
noticed~\cite{karsch:petronzio,gottlieb,kennedy} it can be accurately
measured; being a bulk effect it has a slow volume dependence and most
importantly it is not affected by perturbative contributions which
violate the scaling.
We determine the critical coupling of the deconfinement transition in
lattices with $N_\tau = 3,\,4,\,5,\,6$, and study the pattern of
asymptotic scaling violation of $T_c$, in comparison with the existing
Wilson action simulations (for good reviews
see~\cite{petersson,fingberg:heller:karsch} and references therein),
both in the ``bare'' scheme and in the effective schemes based on the
expectation values of plaquettes. In the bare scheme we find that the
two lattice formulations give quite similar results, taking into
account the redefinition of the $\Lambda$ scale, which corresponds to
the perturbative prediction whitin $10\%$. This is a confirmation that
the Wilson and Symanzik actions are in the same universality class,
and that the $\beta$ function is not affected, in the ``bare'' scheme,
by the irrelevant operators connecting the two lattice formulations:
on the other hand the results show further that with the currently
accessible lattices the two-loop perturbative approximation is not
sufficient to model the renormalization group flow in the scaling
region. In sec.~(\ref{sec:simul}) we give some details on our
simulation and on the numerical results, discussing the consequences
of an effective coupling parameterization.  In sec.~(\ref{sec:fss}) we
discuss a test of our simulation, aimed to verify the first order
nature of the deconfinement transition through a F.S.S.  analysis on
$N_\tau=3$ lattices.

\section{The simulation}
\label{sec:simul}

\begin{figure}
\vskip 12truecm \includegraphics{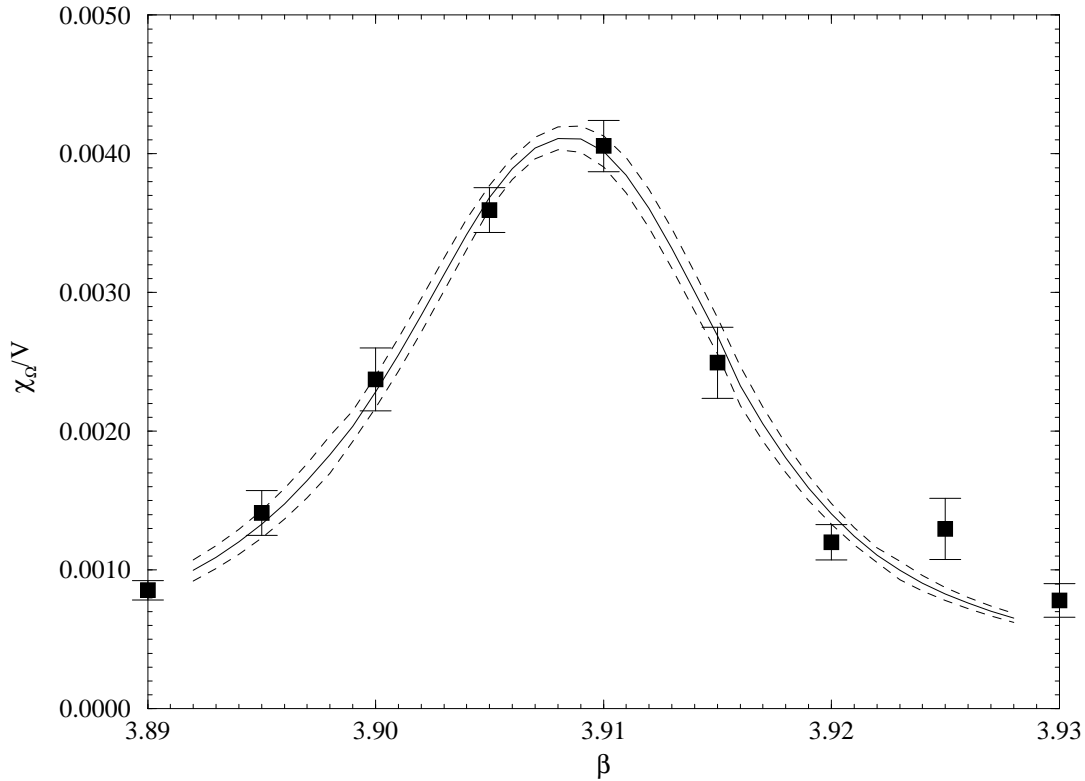}
\caption{Susceptibility of the Polyakov line on a $3\times 12$ lattice}
\label{fig:3x12}
\end{figure}

\begin{figure}
\vskip 11truecm \includegraphics{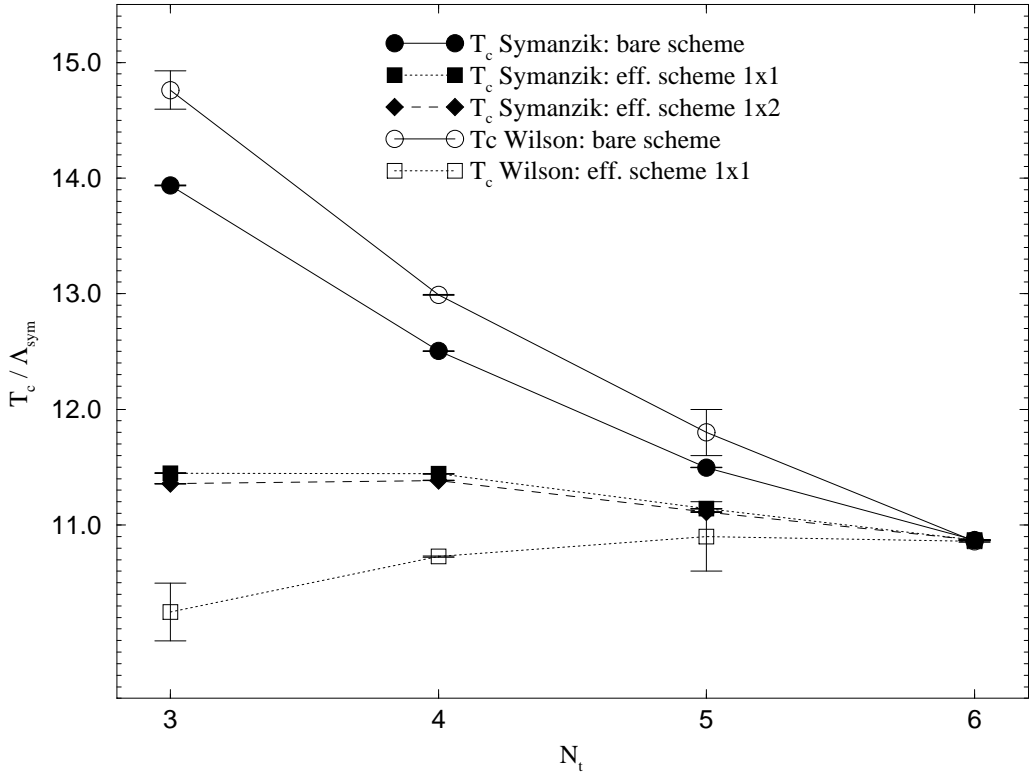}
\caption{Asymptotic scaling figure for the critical temperature}
\label{fig:Tc}
\end{figure}

We have considered the $SU(3)$ gauge theory on asymmetric lattices
using the lowest order tree improved Symanzik action
\begin{equation}
  S_{\rm I} = \beta \sum {\rm Re} {\rm Tr} \left( \frac{5}{3} U_{1
    \times 1} - \frac{1}{12} U_{1 \times 2} \right),
\end{equation}
defined in order to remove $O\left(a^2\right)$ scaling violations in
the Callan-Symanzik equation for physical quantities measured on the
lattice: we refer to our preceding
work~\cite{cella:curci:tripiccio:vicere} and references therein for a
more detailed discussion of the improvement strategies and of the
results existing in literature.  We have performed simulations on
lattices $N_{\tau} \times N_{\sigma}^3 = 3\times 12^3,\,4\times
16^3,\,5\times 20^3,\,6\times 24^3$; the update is done with a
combination of Heat-Bath and of overrelaxation sweeps applied to the
$SU(2)$ subgroups~\cite{cabibbo:marinari}. All the data have been
obtained on the {\sc APE} machines installed in Pisa, including a
$128$ processors machine with a peak speed of 6 GigaFlops. We have
measured the order parameter of the deconfining transition, the
spatial average of the Polyakov line
\begin{equation}
  P = N_{\sigma}^{-3} \sum_{\vec{x}} {\rm Tr}\prod_{t=1}^{N_t}
  U_{4}\left(\vec{x},\,t\right)\ .
\end{equation}
As usual~(see for example~\cite{fukugita:okawa:ukawa}) to consider the
projection of this quantity on the nearest $Z\left(3\right)$ axis,
\begin{equation}
  \Omega = \left\{\begin{array}{ll} {\rm Re} P & \arg{P} \in
  \left(-\frac{\pi}{3},\,\frac{\pi}{3}\right) \cr {\rm Re} P
  \exp\left(-2\,i\,\pi/3\right) & \arg{P} \in
  \left(\frac{\pi}{3},\,\pi\right) \cr {\rm Re}
  P\,\exp\left(2\,i\,\pi/3\right) & \arg{P} \in
  \left(-\pi,\,-\frac{\pi}{3}\right) \cr
\end{array}
\right.\ ,
\end{equation}
in order to avoid the vanishing of the order parameter, on finite
lattices, induced by the tunnelling between broken phases.  We locate
the deconfinement transition by studying the Polyakov susceptibility
\begin{equation}
  \chi\left(\Omega\right) = N_\sigma^3 \left<\left(\Omega\right)^2 -
  \left<\Omega\right>^2\right>
\end{equation}
which exhibits a clear peak structure, and the specific heat of the
lattice action.  The density of states method~\cite{dsm} has been used
to extrapolate data coming from runs at different $\beta$ values and
to combine them in a single curve: the data autocorrelation has been
taken in account using a standard binning procedure, varying the bin
size and testing the existence of a plateau in the variance of the
binned quantities.  In order to reduce the bias induced by the finite
sample size we have used the techniques discussed in~\cite{berg}:
``standard'' and ``jackknife'' estimators have been defined, combining
them with coefficients appropriate to cancel the $\frac{1}{N}$ term in
the bias.  In fig.~(\ref{fig:3x12})\marginpar{\sc Fig.~\ref{fig:3x12}}
is given an example of the
resummation of the Polyakov susceptibility around the deconfining
transition on the $N_\tau = 3$ lattice, where the actual simulation
values are addressed by a black square.
A more extended discussion of the numerical simulation will be given
elsewhere.
In tab.~(\ref{tab:betac}) we report the values of the (pseudo)
critical coupling on the different lattices, obtained by fitting the
peaks with a gaussian curve, and the corresponding critical
temperature in units of $\Lambda_I$, obtained by means of the two loop
R.G. formula
\begin{equation}
  \frac{T_c}{\Lambda_I} = \frac{1}{a\,N_\tau\,\Lambda_I} =
  \frac{1}{N_\tau}\left(\frac{\beta}{2 C_a b_0}\right)^{-b_1/{2
      b_0^2}} \exp\left(\frac{\beta}{4 C_a b_0}\right)\ .
\end{equation}
They are also reported the corresponding quantities obtained in
simulations using the Wilson
action~\cite{karsch:petronzio,gottlieb,kennedy,iwasaki}.  In the last
column the corresponding ratio of the $\Lambda$ parameters in the
improved and standard formulation is listed, showing at $N_\tau=6$ a
10\% discrepancy from the perturbative prediction~\cite{weisz:wohlert}
\begin{equation}
  \frac{\Lambda_I}{\Lambda_W} = 5.29210(1)\ .
\end{equation}
\begin{table}
\caption{$\beta_c$ values on different lattices}
\begin{tabular}{c|c|c|c|c|c}
\hline $N_\tau$ & $\beta_c\left({\rm Improved}\right)$ &
$T_c/\Lambda_I$ & $\beta_c\left({\rm Wilson}\right)$ &
$T_c/\Lambda_W$ & $\Lambda_I / \Lambda_W$\\ 
$3$ & 3.90812(7)  &  13.937(2) & 5.55(1)   & 85.70(96) & 6.39(7) \\ 
$4$ & 4.07252(13) &  12.505(2) & 5.6925(2) & 75.41(2)  & 6.030(2) \\ 
$5$ & 4.19963(14) &  11.497(2) & 5.7933(3) & 68.5(1)  & 5.95(1)\\ 
$6$ & 4.31466(24) &  10.870(3) & 5.8941(5) & 63.05(4)  & 5.800(5)\\ 
\end{tabular}
\label{tab:betac}
\end{table}
It is instructive to compare different actions and schemes by
normalizing the results in order to have the values at smaller $a$
coincide, as in fig.~(\ref{fig:Tc}).\marginpar{\sc Fig.~\ref{fig:Tc}}
It is quite apparent that the asymptotic scaling violations for the
Symanzik and Wilson action, in the ``bare'' scheme, are similar, as
already noted in the $SU(2)$
simulation~\cite{cella:curci:tripiccio:vicere}.
In the same figure we show the same data parameterized in terms of the
expectation values of plaquettes, in the manner of
Parisi~\cite{effective}: to define effective couplings, the
perturbative predictions on the $1\times 1$ and $1\times 2$
plaquettes~\cite{weisz:wohlert} have been used, together with the
results of simulations on large symmetric lattices.  As already
noted~\cite{fingberg:heller:karsch} for the results obtained with the
Wilson action, this parameterization shows an apparent asymptotic
scaling already for $N_\tau \geq 4$; our simulation with the Symanzik
tree improved action shows an apparent scaling already for $N_\tau = 3$.

We recall that the Symanzik approach is based on perturbation theory:
hence the fact that the improvement is apparent when an effective
coupling is used seems a confirmation of the work of Lepage and
Mackenzie~\cite{lepage:mackenzie}, which asserts that the bare
coupling is a bad expansion parameter, and that the perturbative
series, when properly reorganized, work well already on the accessible
lattices.  In other words the effect of the Symanzik improvement is
``buried'' by the large renormalizations induced by the lattice
formulation: the improvement shows up when a scheme is chosen which
minimizes these finite corrections.

\section{Finite size study for $N_\tau = 3$}
\label{sec:fss}

In order to control at least in part the systematic error connected to
the finite size of the lattices, we have performed for $N_\tau=3$ an
additional set of simulations with $N_\sigma =
6,\,7,\,8,\,9,\,10,\,11$, to test the dependence of the critical
coupling on the volume of the system, and to check the expected
scaling relations for a first order phase transition.
\begin{figure}
\vskip 11truecm \includegraphics{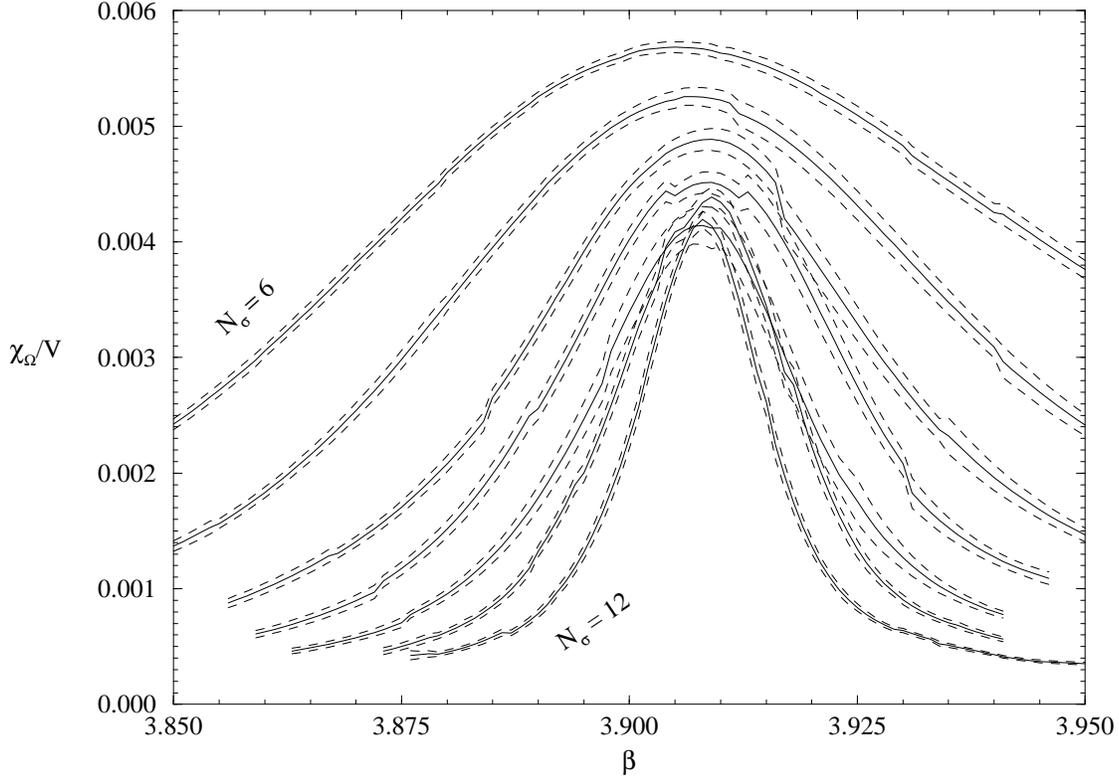}
\caption{Polyakov susceptibility for $N_\tau=3,\,N_\sigma=6,\dots 12$}
\label{fig:3xV}
\end{figure}
We report in tab.~(\ref{tab:fss}) the values of the position of the
peak of the susceptibility of the Polyakov line and its height,
resulting from a gaussian fit with a self consistent data windowing
around the peak.
\begin{table}
\caption{Results of the finite size study on the $N_\tau=3$ lattices.}
\begin{tabular}{c|c|c}
\hline 
$N_\sigma$ & $\beta_c^{\mathrm{max}}$ &
$\chi\left(\Omega\right)^{\mathrm{max}}/N_\sigma^3$ \\
\hline 
  $6$  & $3.90588(34)$  &  $0.00568(5)$ \\ 
  $7$  & $3.90525(53)$  &  $0.00514(12)$ \\ 
  $8$  & $3.90825(46)$  &  $0.00505(12)$ \\ 
  $9$  & $3.90859(24)$  &  $0.00450(12)$ \\ 
  $10$ & $3.90765(5)$   &  $0.00415(16)$ \\ 
  $11$ & $3.90853(8)$   &  $0.00436(9)$ \\ 
  $12$ & $3.90812(7)$   &  $0.00416(7)$ \\ 
\end{tabular}
\label{tab:fss}
\end{table}
In fig.~(\ref{fig:3xV})\marginpar{\sc Fig.~\ref{fig:3xV}}
we compare the results for the Polyakov line
susceptibility: we note that the scaling of the peak is compatible
with the characteristic behavior in presence of a first order phase
transition
\begin{equation}
  \chi\left(\Omega\right) \simeq N_\sigma^3\ .
\end{equation}
Most importantly for our purposes, we note that the shift of the
pseudo-critical coupling corresponding to the peak of the transition
is modest, giving confidence in the values and the errors quoted in
tab.~(\ref{tab:betac}) for the lattices with $N_\tau > 3$.

\section{Conclusions}

In this work we have performed a study of the deconfinement phase
transition in the pure gauge $SU(3)$ theory, using the Symanzik tree
improved action and determining the critical couplings on lattices
with $N_\tau = 3,\,4,\,5,\,6$.  The main conclusion is that, as in the
case of the deconfinement transition in $SU(2)$
theory~\cite{cella:curci:tripiccio:vicere}, the same pattern of
asymptotic scaling violation is observed.  The asymptotic scaling
violations are much reduced if the two loop perturbative expression is
used in conjunction with effective couplings, based on UV dominated
quantities like the plaquette expectation values, and comparison with
the results of simulations with the Wilson action, in this
``effective'' schemes, shows a slighlty precocious scaling, which is
tempting to interpret as an effect of the improvement.  Although we do
not have under control the systematical errors due to the finite
volume, the analysis of the finite size behavior in the $N_\tau=3$
lattices makes us confident that the quoted results are not much
affected by the finite size shift of the transition.
\begin{ack}
  We wish to thank R.~Tripiccione for many useful discussions and for
  a careful reading of the manuscript. We also acknowledge the support
  of the {\sc APE} group, particularly S.~Cabasino and G.~M.~Todesco.
\end{ack}

\end{document}